\documentclass[%
preprint,
superscriptaddress,
amsmath,amssymb,
aps,
]{revtex4-1}
\usepackage{amsmath}
\usepackage{graphicx}
\usepackage{hyperref}
\usepackage{siunitx}
\usepackage{lineno}
\usepackage{multirow} 
\usepackage[version=4]{mhchem}

\usepackage{graphicx}
\usepackage{dcolumn}
\usepackage{bm}
\usepackage{verbatim}   
\usepackage{color}
\usepackage[normalem]{ulem}
\newcommand{\new}[1]{\textcolor{black}{#1}}

\begin{document}

\title{Potential energy curves and transition dipole moments of I$_2$, I$_2^+$ and I$_2^-$: \new{visible} absorption spectroscopy.}

\author{Sylvain~Badin}
\affiliation{
	Universit\'{e} de Lille, CNRS, UMR 8523 - PhLAM - Physique des Lasers, Atomes et Molécules, F-59000 Lille,  France
}
\affiliation{
	Sorbonne Universit\'{e}, CNRS, Laboratoire de Chimie Physique Mati\`{e}re et Rayonnement, UMR 7614, F-75005 Paris, France
}

\author{Andre~Severo~Pereira~Gomes}
\affiliation{
Universit\'{e} de Lille, CNRS, UMR 8523 - PhLAM - Physique des Lasers, Atomes et Molécules, F-59000 Lille,  France
}

\author{Nicolas Sisourat}
\affiliation{
	Sorbonne Universit\'{e}, CNRS, Laboratoire de Chimie Physique Mati\`{e}re et Rayonnement, UMR 7614, F-75005 Paris, France
}
\email{Nicolas.Sisourat@sorbonne-universite.fr}

\date{\today}

\begin{abstract}  
We present the potential energy curves for the ground and lower electronically excited states, as well as transition dipole moments for the I$_2$, I$_2^+$ and I$_2^-$ systems, obtained from relativistic multireference configuration interaction calculations which include spin-orbit coupling effects already at the mean-field description. From these data, we simulated the visible absorption spectra of the three molecules for temperatures ranging from 300K to 1200K. These results serve as basis to a better understanding of the physical and chemical properties of low-pressure iodine plasmas.
\end{abstract}

\maketitle

\section{Introduction}\label{sec.intro}

Electric propulsion is a method of choice for low-power space propulsion, which is used in satellites and spacecraft~\cite{mazouffre_electric_2016, levchenko_recent_2018}. To achieve efficient propulsion, the propellant should have a high atomic/molecular mass and a low ionization energy. These criteria led to the adoption of xenon as the most used fuel for electric propulsion~\cite{croes_effect_2018}. Furthermore, its status of noble gas gives it very low reactivity which allows it to be stored under pressure without particular constraints~\cite{wiley-vch_noble_2001} . However, a major limitation to the use of xenon is its low availability, its high cost and its storage in high-pressure tanks, significantly limiting the usable volume in satellites~\cite{mazouffre_electric_2016}. Thus, it appears necessary to develop the use of propellant alternatives.

In this context, iodine is a promising candidate~\cite{grondein_global_2016}. However, the basic physics and chemistry of low-pressure iodine plasmas are not well-understood~\cite{dietz_molecular_2019}. Recent experimental and theoretical efforts have therefore been made to bridge this gap~\cite{esteves_charged-particles_2022,esteves_two-photon_2023, poline_final-state-resolved_2022, badin_theoretical_2023}. In order to compare experiments and theory, accurate diagnostic techniques are necessary. Absorption spectroscopy is a powerful tool in this context since it can provide the concentration and temperature of the species present in the plasma. However, such diagnostic technique relies on the knowledge of accurate absorption spectra of the light-absorbing species.

This work is devoted to the calculation of the visible absorption spectra of I$_2$, I$_2^+$ and I$_2^-$ at different temperatures (T). We first present the potential energy curves of I$_2$ \new{(section III.A)}, I$_2^+$ \new{(section III.B)} and I$_2^-$ \new{(section III.C)} and we compare the spectroscopic constants resulting from these curves with the available experimental data. The dipole transition moments between the ground states and the dipole-allowed excited states are then reported. The absorption spectra obtained from these potential energy curves and transition moments are finally discussed. 

The visible absorption spectra of I$_2$ has been extensively investigated~\cite{esteves_measurement_2024} at T=300K and thus serves as a benchmark for our computational approach. However, in a plasma the temperature of the ions can be higher. We therefore provide the spectra up to T=1200K. \new{Much less is known for  I$_2^+$ and I$_2^-$. Therefore, the results reported in this study bridge an important gap.}
 
\section{Computational details}\label{sec.th}

The potential energy curves of I$_2^+$ and I$_2^-$ were calculated using the
4-component multi-reference configuration interaction (MRCI) method, as implemented in the KRCI module~\cite{fleig_generalized_2003} of the DIRAC program~\cite{saue_dirac_2020, DIRAC25}. A fully uncontracted triple-zeta basis set including a total of three diffuse functions~\cite{dyall_relativistic_2006} for each angular momentum was used. We note this corresponds to adding two additional diffuse primitives than those available for the standard basis sets, with the exponents being determined as an even-tempered sequence from the two most diffuse exponents in the standard basis set. The active space contains the 12 spinors formed from the 5p orbitals of iodine (in order of increasing energy $\sigma_{1/2}, \pi_{1/2}, \pi_{3/2}, \pi^{*}_{3/2}, \pi^{*}_{1/2}$ and $\sigma^{*}_{1/2}$), which are occupied by 9, 10 and 11 electrons for I$_2^+$,  I$_2$ and I$_2^-$, respectively. For each system the potential energy curves were calculated for 34 geometries corresponding to internuclear distances varying between 2 and 8 \AA. The dipole transition moments between these states were also obtained with the same approach. The potential energy curves of I$_2$ are the same curves as reported in ref.~\cite{poline_final-state-resolved_2022}, and were also obtained at same level of theory as described above.

We also carried out calculations with equation of motion coupled cluster singles and doubles approach (\new{labeled EOMCC hereafter}) with the RELCCSD module of the DIRAC program, to serve as cross-validation for the MRCI potential energy curves of the low-lying states of I$_2$. Since RELCCSD does not currently implement the calculation of transition dipole moments for any of the EOMCC models, we were unable to compare the respective MRCI and EOMCC absorption spectra.

\new{The absorption spectra were computed as follows: The first step is to compute the fundamental vibrational wavefunction in electronic state $i$. For this, we use the relaxation-by-propagation method with imaginary time \cite{desouterlecomte2017}. The time propagation of the wavepacket is carried out using the MCTDH package \cite{meyermultidimensional2009}.}

\new{Assuming the sudden approximation \cite{desouterlecomte2017}, the
initial wavefunction is multiplied by the corresponding dipole transition moment and is placed on the potential energy curve of state $f$.  Next, the promoted wavepacket is propagated numerically (also using the MCTDH package) on the potential energy curve of state $f$. The absorption spectrum can then be obtained from the Fourier transform of the autocorrelation function of the promoted packet:
\begin{equation}
	\sigma\left(\omega\right)=\frac{2\pi\omega}{\hbar c}\int_{-\infty}^{+\infty}C\left(t\right)\exp\left(i\left(\frac{E_{0}}{\hbar}+\omega\right)t\right)dt
\end{equation}
where $E_{0}$ is the energy of the initial state and $C\left(t\right)$ is the
autocorrelation function of the promoted packet:
\begin{equation}
	C\left(t\right)=\left\langle \Psi_{f}\left(t=0\right)\right.\left|\Psi_{f}\left(t\right)\right\rangle.
\end{equation}
}

\new{The total absorption spectrum is the sum of all the absorption spectra originating from each vibrational state $\nu$ of $i$, weighted by their respective Boltzmann populations at the temperature considered.  We consider hereafter the transitions from the first 16 vibrational states of the ground electronic state since the states with higher energy have a \new{total contribution of less than 1\%} at the highest temperature.}

\section{Results}\label{sec.res}

In the following we present the spectroscopic constants for the calculated electronic states of the three species, and the potential energy curves associated to states which contribute to the final absorption spectra.

\subsection{I$_2$}

\begin{figure}
\centering
\includegraphics[width=0.7\linewidth]{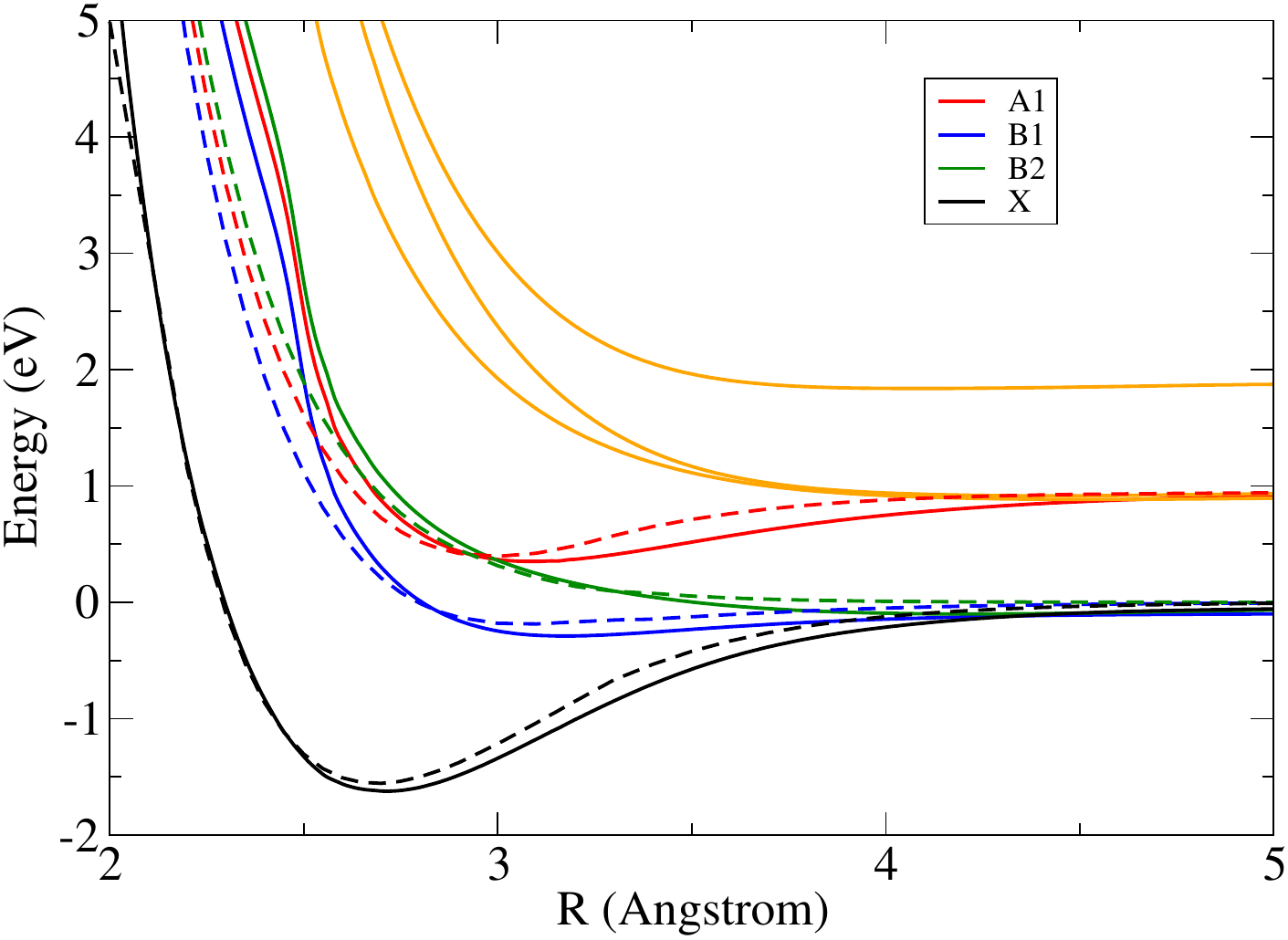} \\
\includegraphics[angle=0, width=0.7\linewidth]{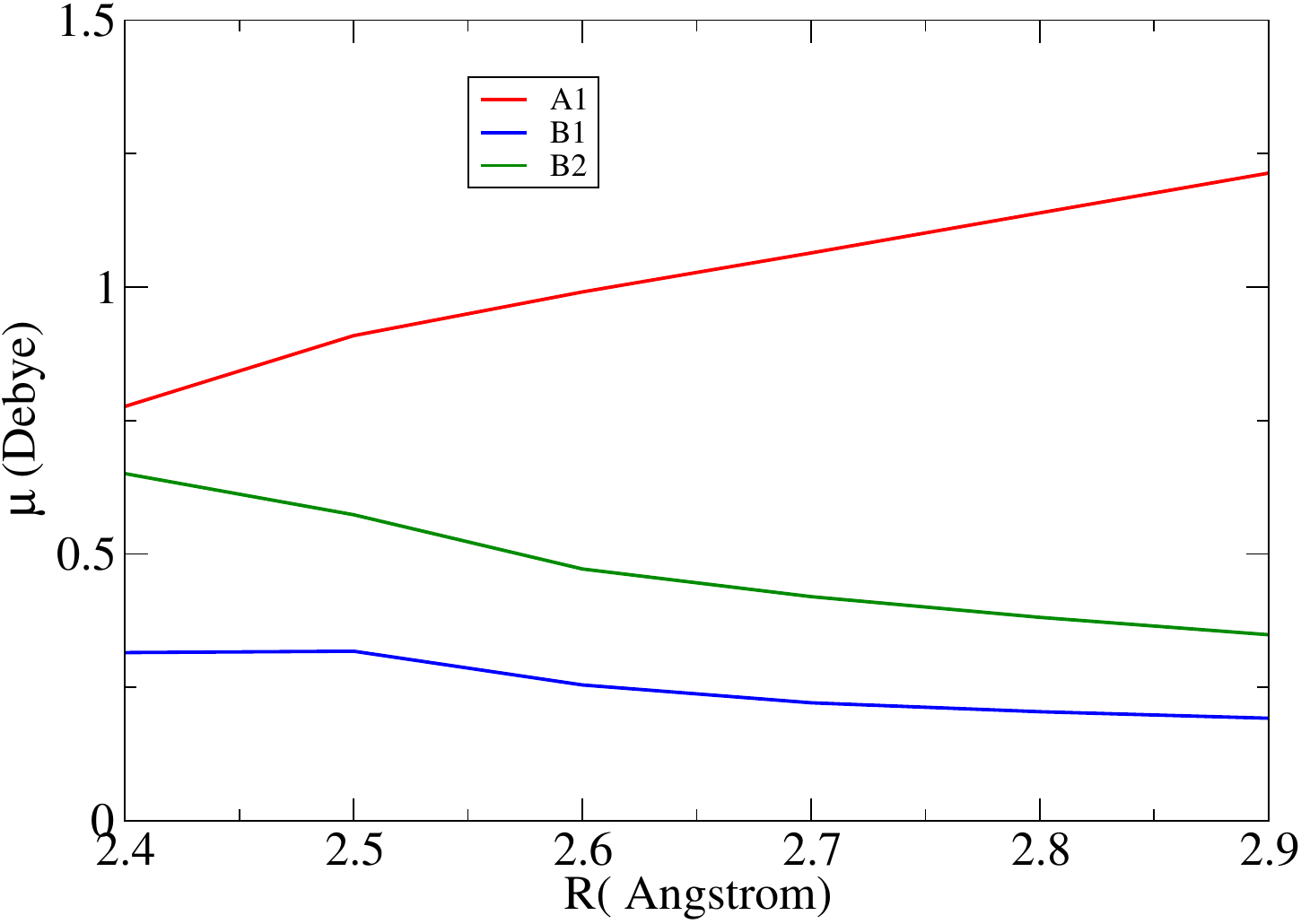}
\caption{Top\label{fig:pecs_I2} MRCI (solid lines) and EOMCC (dashed lines) potential energy curves for the I$_2$ molecule (in eV) for different internuclear distances (R, in \AA), for the X ground state (black), the first B1 state (blue), the first B2 state (green), and the first A1 (red) state. The orange lines represent higher energy states that do not contribute to the absorption spectra shown below. Bottom:\label{fig:tdm_I2}  MRCI transition dipole moments (in Debye) of I$_2$ between the ground and B1 (blue), B2 (green) and A1 (red) excited states near the ground-state equilibrium structure.}
\end{figure}

The MRCI (calculated previously by some of us~\cite{poline_final-state-resolved_2022}) and EOMCC potential energy curves of
I$_2$ are shown in Figure~\ref{fig:pecs_I2}, whereas a comparison of their spectroscopic constants with those measured by~\cite{lukashov_iodine_2018} is given in table~\ref{tab:comp_MRCI_EOMCC_I2}. 

From these results we observe a qualitative agreement between the curves obtained by the two methods, but with some notable differences. Particularly, the repulsive part of the curves MRCI is shifted by around 0.2 \AA\ towards larger internuclear distances compared to the EOMCC curves. The spectroscopic constants $T_e$ and $D_e$ obtained with EOMCC are in better agreement with the experimental data than those obtained with the MRCI method. Conversely, the EOMCC harmonicity constants ($\omega_0$) for the excited states show worse agreement to experiment than the MRCI ones, with the exception of that of the ground state which is within 0.5 cm$^{-1}$ of the experimentally derived value. EOMCC also typically overestimates the values for anharmonic constants ($\omega_e x_e$) whereas MRCI tends to underestimate them with respect to experiment.

We observe that MRCI predicts a weakly bound B2 state (green curve in Figure~\ref{fig:pecs_I2}), in agreement with the experimental data from
Lukashov \textit{et al.}\cite{lukashov_iodine_2018} that does indicate it to be  very weakly bound. The EOMCC potential energy curves, on the other hand, point to a dissociative state, a behavior which might be due to higher order correlation or relaxation contributions which are lacking in the underlying \new{EOMCC} wavefunction and the restriction to singles and doubles excitations in EOMCC~\cite{shee_equation--motion_2018-1,pototschnig_electronic_2021-1}. An investigation of these higher-lying states with EOMCC including higher-order excitations would be useful to better understand the shortcomings of the current calculations, but is beyond the scope of this work since higher-order EOMCC wavefunctions are currently not implemented in the DIRAC code.

The transition dipole moments for the allowed transitions of I$_2$ in the region around the ground-state equilibrium distance are shown in Fig.~\ref{fig:tdm_I2}. We consider only the transitions towards the first state of symmetry ($\Omega$ = 1$_u$) and the first two symmetry states ($\Omega$ = 0$_u^+$). Dipole-allowed transitions to others states are too high in energy to absorb in the visible. \new{From it, we observe a slight decrease for the transition towards the B1 and B2 states as the internuclear distance increases, whereas a slight increase is seen for the transition towards the A1 state.}

\begin{table}
\caption{Spectroscopic constants of X, A1, B1 and B2 states for the I$_{2}$ molecule. \label{tab:comp_MRCI_EOMCC_I2}}
\centering{}%
\begin{tabular}{ccccccccc}
\hline 
Symmetry & State & $\Delta E$ (eV) & $T_{e}$(cm$^{-1}$) & $\omega_{0}$(cm$^{-1}$) & $\omega_{e}x_{e}$(cm$^{-1}$) & $R_{e}$(Å) & $D_{e}$(eV) & \multicolumn{1}{c|}{Method}\tabularnewline
\hline 
$\left(\Omega=0\right)_{g}^{+}$ & {X} & 0 & 0 & 236 & 0.75 & 2.72 & 1.63 & MRCI\tabularnewline
 &  &  &  & 215 & 0.91 & 2.69 & 1.56 & EOMCC\tabularnewline
 &  &  &  & 214.5 & 0.608 & 2.67 & 1.556 & Exp. \cite{lukashov_iodine_2018}\tabularnewline
\hline 
$\left(\Omega=0\right)_{u}^{+}$ & A1 & 2.46 & 1.97 & 125 & 0.297 & 3.09 & 0.58 & MRCI\tabularnewline
 &  & 2.29 & 1.95 & 150 & 1.29 & 2.99 & 0.54 & EOMCC\tabularnewline
 &  &  & 1.955 & 125.7 & 0.753 & 3.02 & 0.543 & Exp. \cite{lukashov_iodine_2018}\tabularnewline
\hline 
$\left(\Omega=1\right)_{u}$ & B1 & 1.87 & 1.38 & 103 & 1.13 & 3.17 & 0.29 & MRCI\tabularnewline
 &  & 1.77 & 1.36 & 133 & 2.89 & 3.06 & 0.19 & EOMCC\tabularnewline
 &  &  & 1.352 & 92.95 & 1.56 & 3.11 & 0.209 & Exp. \cite{lukashov_iodine_2018}\tabularnewline
\hline 
$\left(\Omega=1\right)_{u}$ & B2 & 2.67 & 1.46 & 32.0 & 0.32 & 4.23 & 0.10 & MRCI\tabularnewline
 &  & 2.49 & - & - & - & - & - & EOMCC\tabularnewline
 &  &  &  & 20.8 & 0.51 & 4.45 & 0.031 & Exp. \cite{lukashov_iodine_2018}\tabularnewline
\hline 
\end{tabular}
\end{table}

\begin{figure}
\centering
\includegraphics[width=0.7\linewidth]{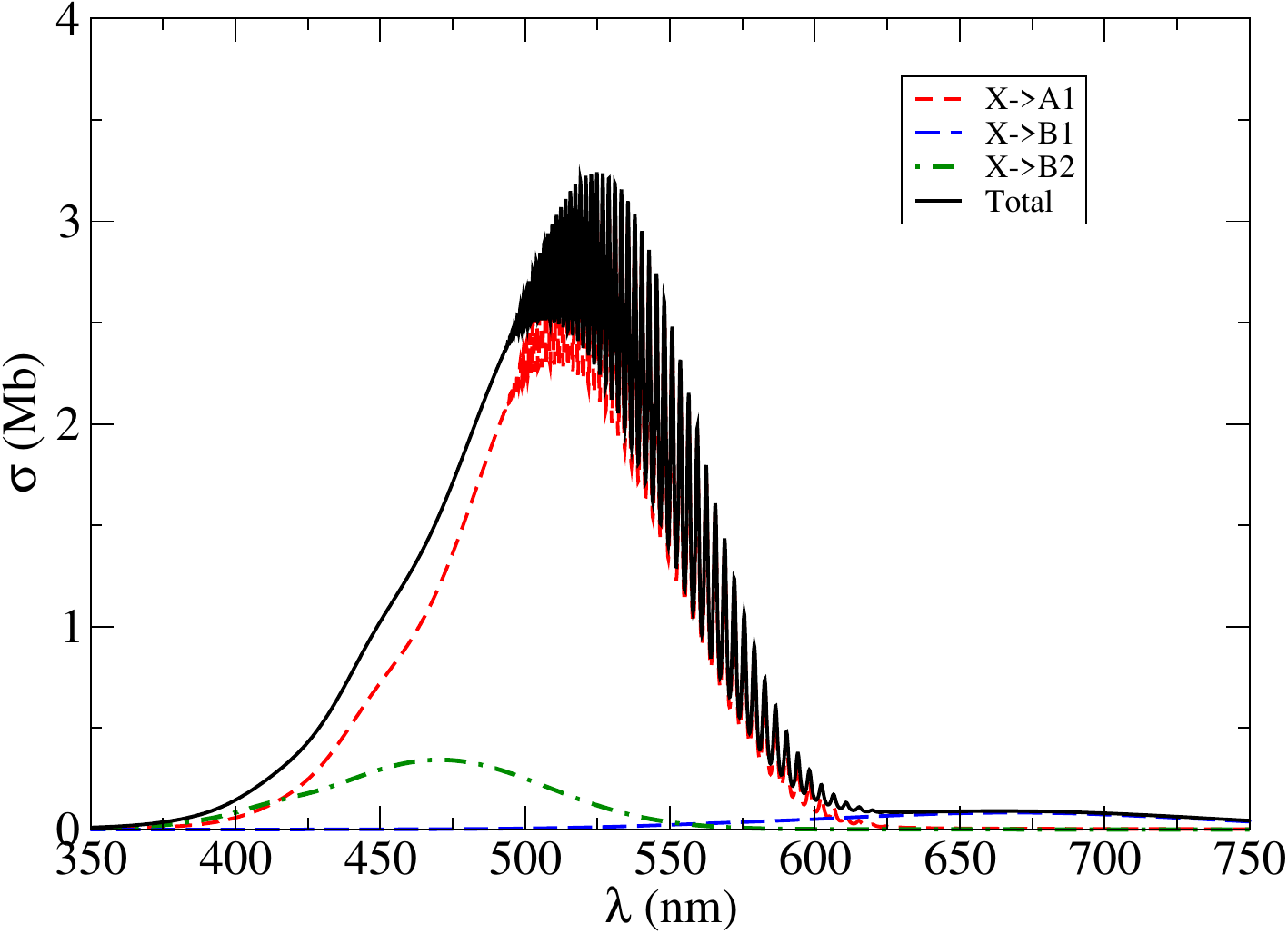}
\includegraphics[width=0.7\linewidth]{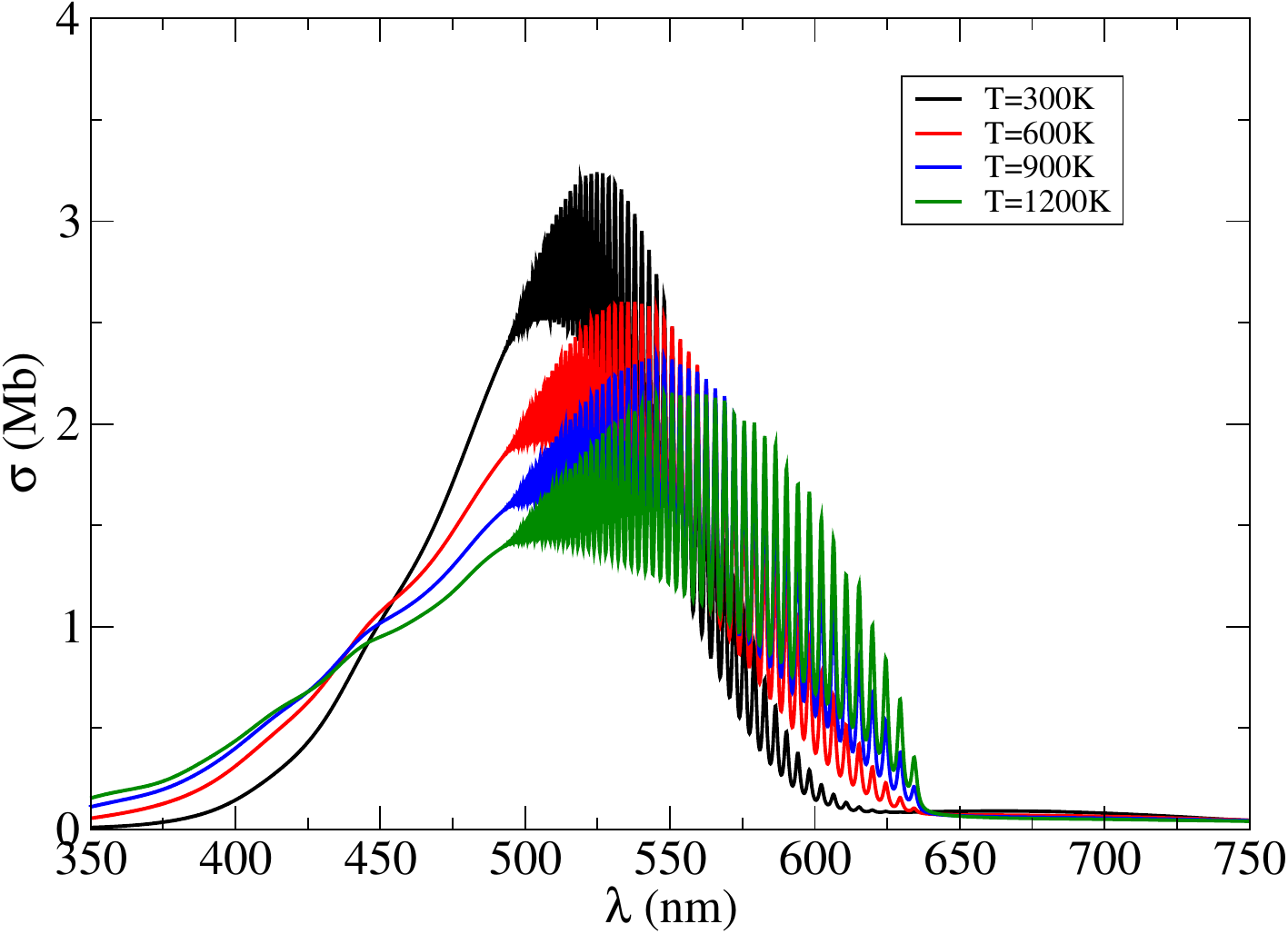}
\caption{Top:\label{fig:specti2det} State-resolved absorption spectrum of I$_2$ at T=300K. Bottom:\label{fig:spectI2_T} Absorption spectrum of I$_2$ at different temperatures. In both cases results are obtained from the MRCI potential energy curves.}
\end{figure}

Figure~\ref{fig:specti2det} shows the absorption spectra towards states A1, B1 and B2 calculated from the MRCI curves at 300K. 
The total absorption spectrum at a temperature T = 300K, towards each of the states, is then obtained by averaging the spectra obtained for each vibrational state, weighted by the population p($\nu_i$) of this state, given by a Boltzmann distribution. We observe the spectrum is dominated by the absorption towards state A1 which forms a broad band centered around 540 nm with an apparent vibrational structure beyond 500 nm. The calculated band is broadened towards short wavelengths, which is consistent with the fact that the MRCI potential energy curves of the excited states seem to overestimate the repulsive region. The absorption towards the B1 and B2 states contribute to the total absorption spectrum via two broad bands without vibrational structure (which is consistent with their weakly bound nature) centered at 690 and 500 nm respectively.  Increasing the temperature leads to a broadening of the spectra and to a shift of their absorption maximum towards long wavelengths, as one can see from the evolution of the absorption spectra as a function of temperature presented in Figure~\ref{fig:spectI2_T}.

\begin{figure}
\centering
\includegraphics[width=0.7\linewidth]{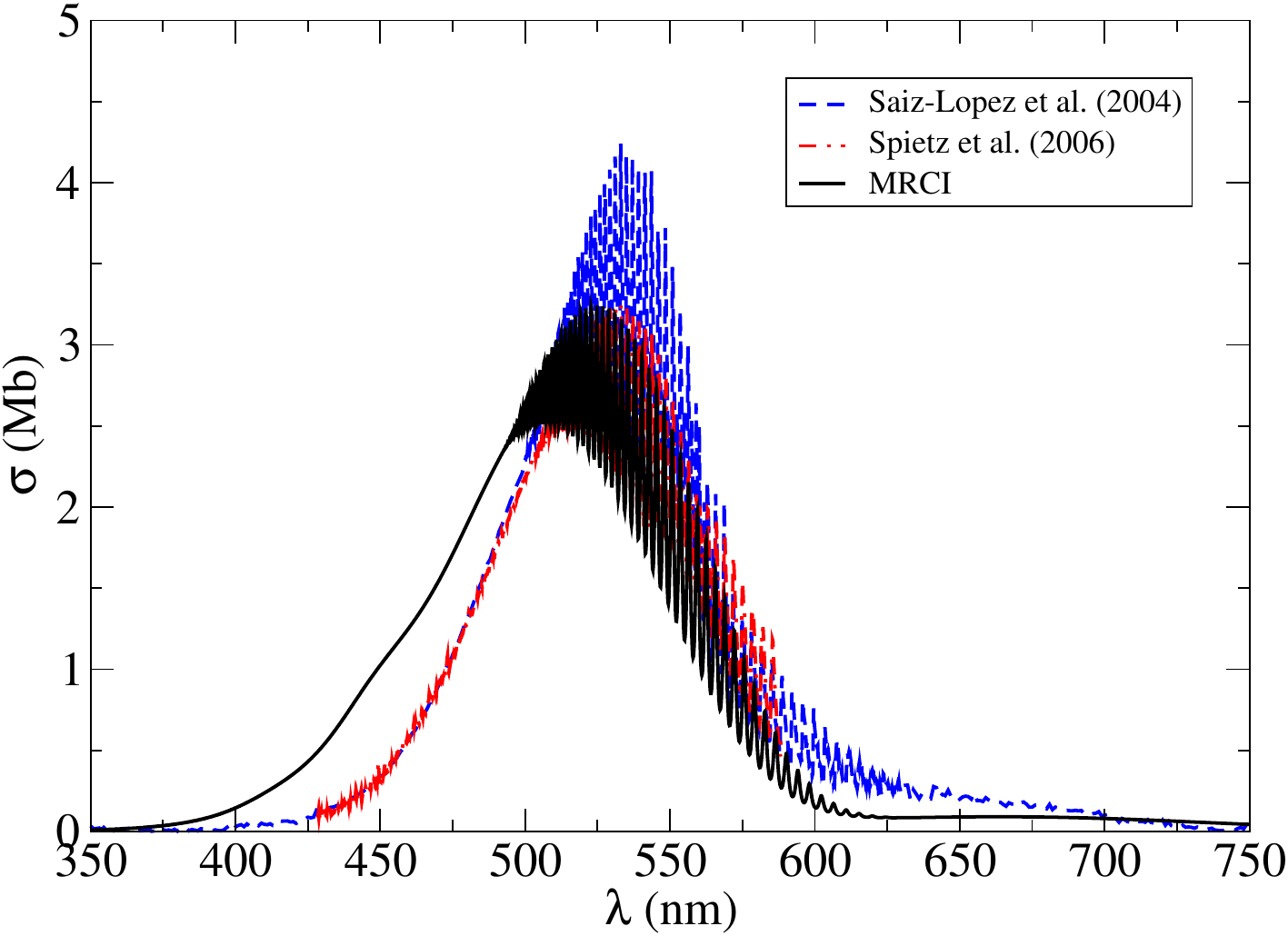}
\caption{\label{fig:specti2exp} Comparison of our calculated absorption at T=300K with those obtained experimentally by Saiz-Lopez~\textit{et al.}~\cite{saiz-lopez_absolute_2004} and Spietz~\textit{et al.}~\cite{spietz_effects_2006}.}
\end{figure}

Figure~\ref{fig:specti2exp} presents a comparison between these spectra and those obtained experimentally by Saiz-Lopez~\textit{et al.}~\cite{saiz-lopez_absolute_2004} and Spietz~\textit{et al.}~\cite{spietz_effects_2006}. The spectrum obtained from the MRCI curves shows a slight  disagreement with the experimental results, particularly between 400 and 500 nm where its absorption is overestimated by up to 0.5 Mb. This indicates that the repulsive part of the potential energy curve for states A1, B1 and B2 is probably overestimated, and that as discussed above the EOMCC description at short range is indeed closer to the experimental picture.

\subsection{I$_2^+$}

\begin{figure}
\centering
\includegraphics[width=0.7\linewidth]{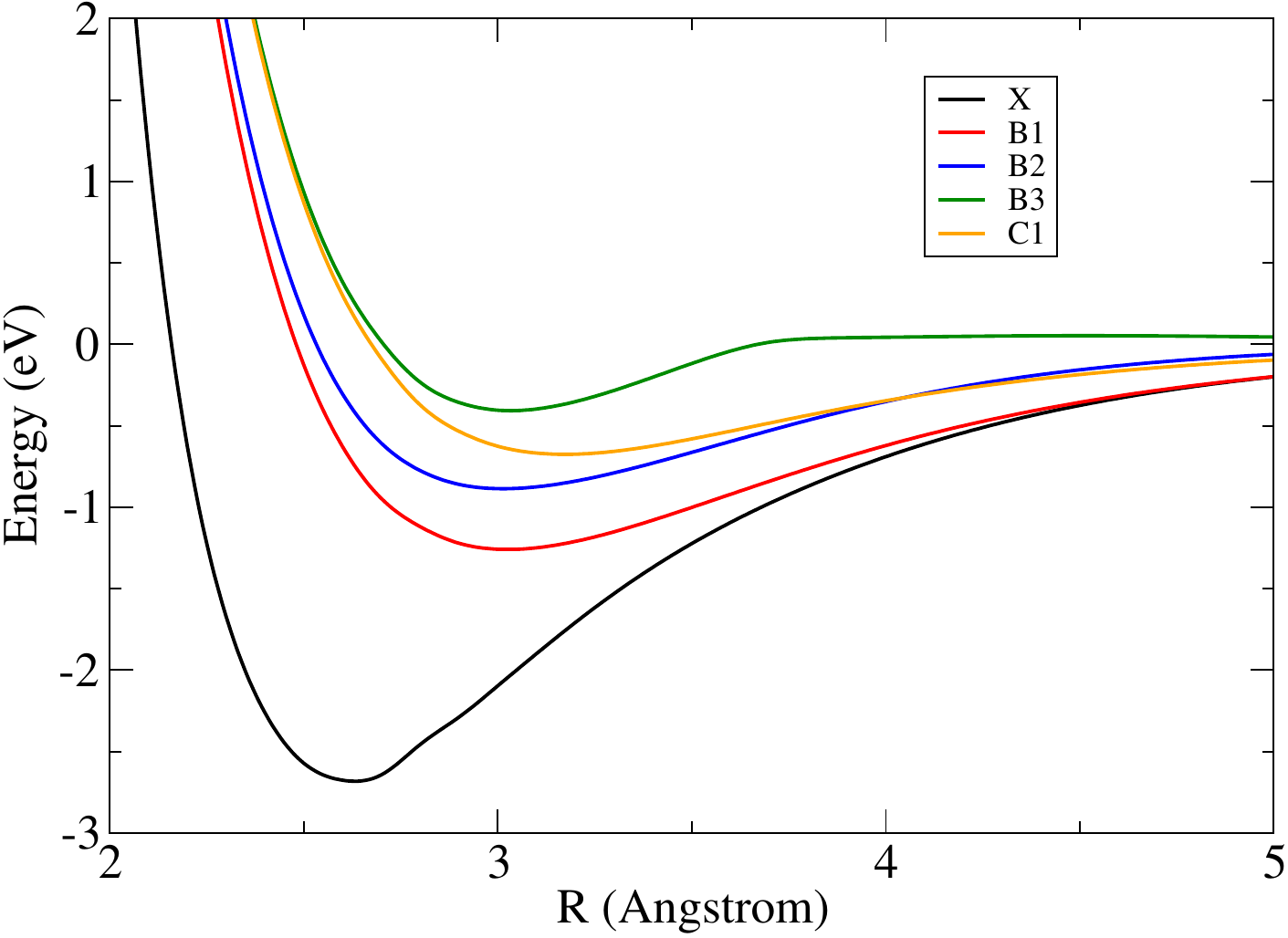}\\ 
\includegraphics[angle=0, width=0.7\linewidth]{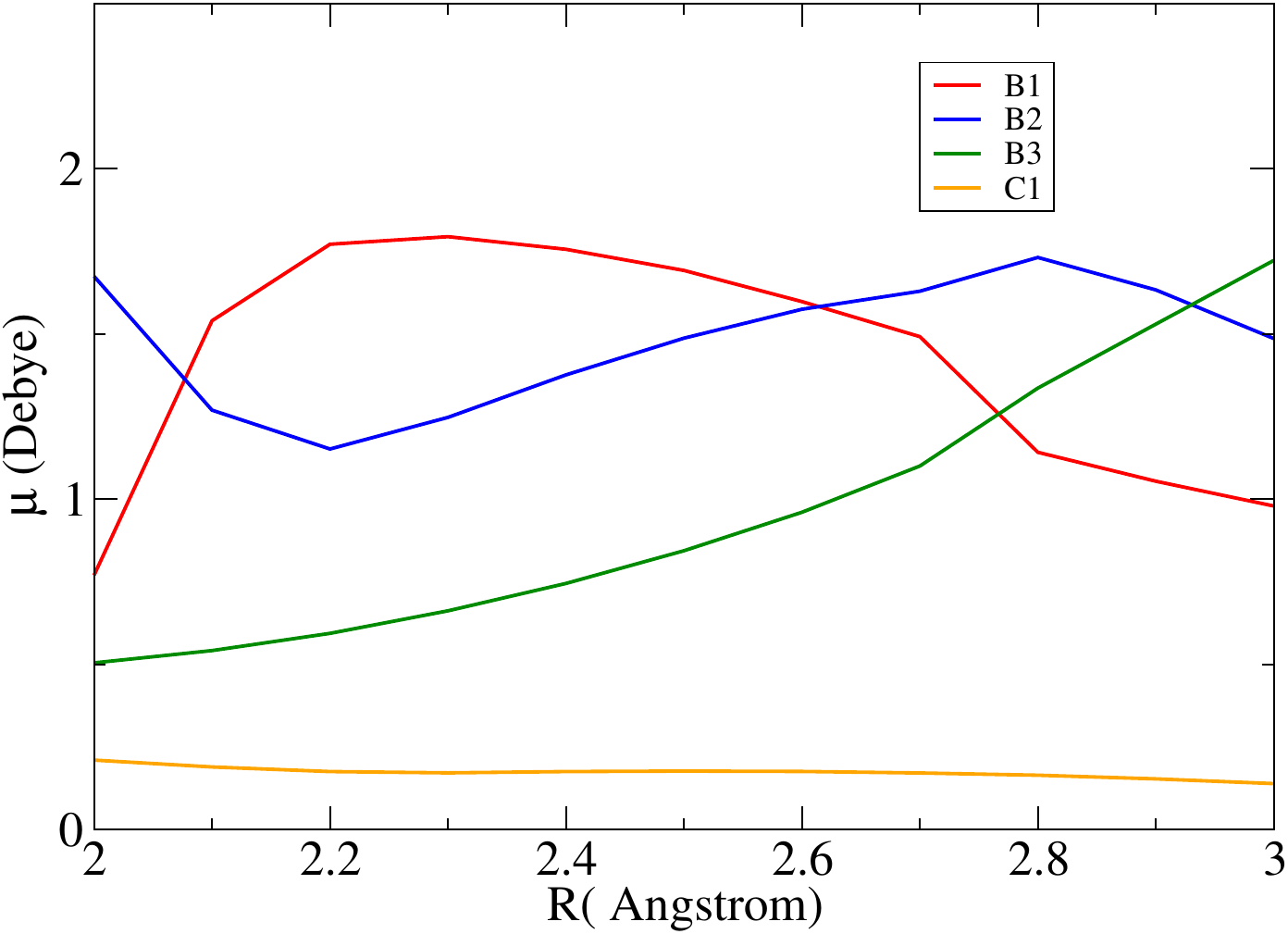}
\caption{Top: MRCI \label{fig:pecs_I2p} potential energy curves (in eV) for the $^{2}X$ ground-state (black) as well as the B1 (red), B2 (blue), B3 (green) and C1 (orange) excited states for different internuclear distances (R, in \AA). Bottom:\label{fig:tdm_I2p} Transition dipole moments (bottom, in Debye)) in the region around the ground-state equilibrium distance between the ground and excited states of I$_2^+$.}
\end{figure}

The potential energy curves of I$_2^+$ are given in Fig.~\ref{fig:pecs_I2p}. In total 26 electronic states have been calculated (5 for the symmetries\footnote{here $\Omega$ denotes the projection of the total electronic angular momentum on the internuclear axis} ($\Omega$ = 1/2$_g$), ($\Omega$ = 1/2$_u$), ($\Omega$ = 3/2$_g$)
and ($\Omega$ = 3/2$_u$) and 3 for the symmetries ($\Omega$ = 5/2$_g$) and ($\Omega$ = 5/2$_u$). The ground state is of symmetry $\Omega$ = 3/2$_g$).

There are two dissociation limits corresponding to these potential energy curves. The first one corresponds to I and I$^+$ in their ground electronic states (i.e. $^2$P$_{3/2}$ and $^3$P$_{2}$, respectively). We set the energy of this dissociation limit to zero. A second dissociation limit, located 0.8 eV above the first one corresponds to the combination of I($^2$P$_{3/2}$) and I$^+$($^3$P$_{1}$), or I($^2$P$_{3/2}$) and I$^+$($^3$P$_{0})$, electronic states of neutral and cationic forms of the iodine atom. The energy difference between these two states of I$^+$ is only 0.07 eV, which is barely visible in the figure.

The potential energy curves of the first symmetry states ($\Omega$ = 3/2$_g$), ($\Omega$ = 3/2$_u$) and ($\Omega$ = 1/2$_g$) reproduce with fairly good agreement the values of the spectroscopic constants measured by Deng~\textit{et al.}~\cite{deng_high-resolution_2012} or Cockett~\textit{et al.}~\cite{cockett_zero_1995,cockett_zero_1996}. As can be seen from Table~\ref{tab:Spectro_I2+}, the differences between the theoretical and experimental equilibrium distances of these states are lower than 0.1 Å and the relative differences for the vibration and anharmonicity constants are less than 5\% and 10\%, respectively.

\begin{table}
\resizebox{0.8\textwidth}{!}{
	\begin{tabular}{cccccccc}
		Symmetry & State & Dissociation limit & $\Delta_{e}$ (eV) & $\omega_{e}$ (cm$^{-1}$) & $\omega_{e}x_{e}$ (cm$^{-1}$) & $R_{e}$ (\r{A}) & $D_{e}$ (eV) \tabularnewline
		\hline 
		\multirow{6}{*}{$\left(\Omega=1/2\right)_{g}$} & \multirow{2}{*}{1} & \multirow{2}{*}{$^{2}\ce{P}_{3/2}+{}^{3}\ce{P}_{2}$} & 0.70 & 239  & 0.911  & 2.6  & 1.95 \tabularnewline
		&  &  &  & (229 \cite{cockett_zero_1996}) & (0.81 \cite{cockett_zero_1996}) & (2.6 \cite{cockett_zero_1996}) & \tabularnewline
		& 2 & $^{2}\ce{P}_{3/2}+{}^{3}\ce{P}_{2}$ & 3.34 & {*} & {*} & {*} & {*}\tabularnewline
		& 3 & $^{2}\ce{P}_{3/2}+{}^{3}\ce{P}_{2}$ & 4.26 & 74.2 & 1.21 & 3.4 & 0.14\tabularnewline
		& 4 & $^{2}\ce{P}_{3/2}+{}^{3}\ce{P}_{2}$ & 4.40 & - & - & - & -\tabularnewline
		& 5 & $^{2}\ce{P}_{3/2}+{}^{3}\ce{P}_{0/1}$ & 4.57 & {*} & {*} & {*} & {*}\tabularnewline
		\hline 
		\multirow{5}{*}{$\left(\Omega=1/2\right)_{u}$} & 1 & $^{2}\ce{P}_{3/2}+{}^{3}\ce{P}_{2}$ & 2.00 & 114 & 0.288 & 3.1 & 1.41\tabularnewline
		& 2 & $^{2}\ce{P}_{3/2}+{}^{3}\ce{P}_{2}$ & 2.86 & 93.2 & 0.428 & 3.2 & 0.63\tabularnewline
		& 3 & $^{2}\ce{P}_{3/2}+{}^{3}\ce{P}_{2}$ & 3.03 & 124 & 1.89 & 3.0 & 0.25\tabularnewline
		& 4 & $^{2}\ce{P}_{3/2}+{}^{3}\ce{P}_{2}$ & 3.80 & 126 & {*} & 3.2 & {*}\tabularnewline
		& 5 & $^{2}\ce{P}_{3/2}+{}^{3}\ce{P}_{0/1}$ & 5.71 & - & - & - & -\tabularnewline
		\hline 
		\multirow{6}{*}{$\left(\Omega=3/2\right)_{g}$} & \multirow{2}{*}{1} & \multirow{2}{*}{$^{2}\ce{P}_{3/2}+{}^{3}\ce{P}_{2}$} & 0 & 246 & 0.698 & 2.6  & 2.69\tabularnewline
		&  &  &  & (239 \cite{deng_high-resolution_2012}) & (0.650 \cite{deng_high-resolution_2012}) & (2.58 \cite{deng_high-resolution_2012}) & \tabularnewline
		& 2 & $^{2}\ce{P}_{3/2}+{}^{3}\ce{P}_{2}$ & 3.91 & 113 & 0.720 & 3.3 & 0.55\tabularnewline
		& 3 & $^{2}\ce{P}_{3/2}+{}^{3}\ce{P}_{2}$ & 3.99 & 139 & {*} & 3.2 & {*}\tabularnewline
		& 4 & $^{2}\ce{P}_{3/2}+{}^{3}\ce{P}_{0/1}$ & 5.38 & {*} & {*} & {*} & {*}\tabularnewline
		& 5 & $^{2}\ce{P}_{3/2}+{}^{3}\ce{P}_{0/1}$ & 5.47 & {*} & {*} & {*} & {*}\tabularnewline
		\hline 
		\multirow{6}{*}{$\left(\Omega=3/2\right)_{u}$} & \multirow{2}{*}{1} & \multirow{2}{*}{$^{2}\ce{P}_{3/2}+{}^{3}\ce{P}_{2}$} & 2.05 & 133 & 0.426  & 3.0  & 1.28\tabularnewline
		&  &  &  & (138 \cite{deng_high-resolution_2012}) & (0.450 \cite{deng_high-resolution_2012}) & (2.95 \cite{deng_high-resolution_2012}) & \tabularnewline
		& 2 & $^{2}\ce{P}_{3/2}+{}^{3}\ce{P}_{2}$ & 2.37 & 126 & 0.541 & 3.0 & 0.91\tabularnewline
		& 3 & $^{2}\ce{P}_{3/2}+{}^{3}\ce{P}_{2}$ & 3.06 & 159 & {*} & 3.0 & {*}\tabularnewline
		& 4 & $^{2}\ce{P}_{3/2}+{}^{3}\ce{P}_{0/1}$ & 5.49 & {*} & {*} & {*} & {*}\tabularnewline
		& 5 & $^{2}\ce{P}_{3/2}+{}^{3}\ce{P}_{0/1}$ & 6.54 & 68.2 & 0.320 & 3.8 & 0.45\tabularnewline
		\hline 
		\multirow{3}{*}{$\left(\Omega=5/2\right)_{g}$} & 1 & $^{2}\ce{P}_{3/2}+{}^{3}\ce{P}_{2}$ & 3.45 & 111 & 0.388 & 3.3 & 0.99\tabularnewline
		& 2 & $^{2}\ce{P}_{3/2}+{}^{3}\ce{P}_{2}$ & 4.47 & 59.7 & 0.789 & 3.5 & 0.14\tabularnewline
		& 3 & $^{2}\ce{P}_{3/2}+{}^{3}\ce{P}_{0/1}$ & 6.79 & 52.6 & 0.260 & 4.2 & 0.33\tabularnewline
		\hline 
		\multirow{3}{*}{$\left(\Omega=5/2\right)_{u}$} & 1 & $^{2}\ce{P}_{3/2}+{}^{3}\ce{P}_{2}$ & 2.98 & 101 & 0.460 & 3.2 & 0.69\tabularnewline
		& 2 & $^{2}\ce{P}_{3/2}+{}^{3}\ce{P}_{2}$ & 5.36 & - & - & - & -\tabularnewline
		& 3 & $^{2}\ce{P}_{3/2}+{}^{3}\ce{P}_{0/1}$ & 6.66 & 81.7 & 0.399 & 3.6 & 0.52\tabularnewline
		\hline 
	\end{tabular}}
	\caption{Spectroscopic constants for $\ce{I2+}$. The values in
		parentheses correspond to experimental data. States for which a
		``-'' replaces the spectroscopic values are dissociative states. The
		presence of a ``{*}'' in place of $\omega_{e}$ and $R_{e}$ means
		that an avoided crossing does not allow the minimum of the state to
		be defined unambiguously.}
	\label{tab:Spectro_I2+}
\end{table}

The ground state X of I$_2^+$ being of symmetry $\Omega$ = 3/2$_g$, the dipole selection rules indicate that the only transitions allowed are towards the symmetry states $\Omega$ = 1/2$_u$,
$\Omega$ = 3/2$_u$ and $\Omega$ = 5/2$_u$. However, the dipole transition moments between states
$\Omega$ = 3/2$_g$ and $\Omega$ = 1/2$_u$ are negligible (i.e. less than 10$^{-7}$ Debye). One can rationalize this by comparing states resulting from a relativistic calculation such as these obtained in this work with those resulting from non-relativistic calculations as done by De Jong~\textit{et al.}~\cite{de_jong_relativistic_1997}. According to these, the ground state is dominated by contributions from a  $^2\Pi_g$ spin-free state while the first state of symmetry $\Omega$ = 1/2$_u$ mainly comes from a $^4\Sigma_u^-$ state. The transition between these two states therefore violates the spin conservation rule, which can explains that the dipole transition moments between these two states are negligible. The other states of symmetry $\Omega$ = 1/2$_u$ are not composed of a single spin-free configuration and there is no obvious explanation for the fact
that their dipole transition moments with the ground state are negligible. Finally, only the transitions from the ground state to the first three 
states of symmetry $\Omega$ = 3/2$_u$ (denoted B1, B2 and B3) and the first state of symmetry $\Omega$ =5/2$_u$(denoted C1) correspond to transitions in the visible, with non-zero dipole transition moments. The dipole transition moments associated with these four transitions are given in Fig.~\ref{fig:tdm_I2p} for the region around the ground-state equilibrium geometry. From them, we observe somewhat more 
variations in the magnitude of the transition dipole moments with the change in  internuclear distances than for I$_2$, though these also remain fairly constant in the range of distances investigated.

\begin{figure}
\centering
\includegraphics[width=0.7\linewidth]{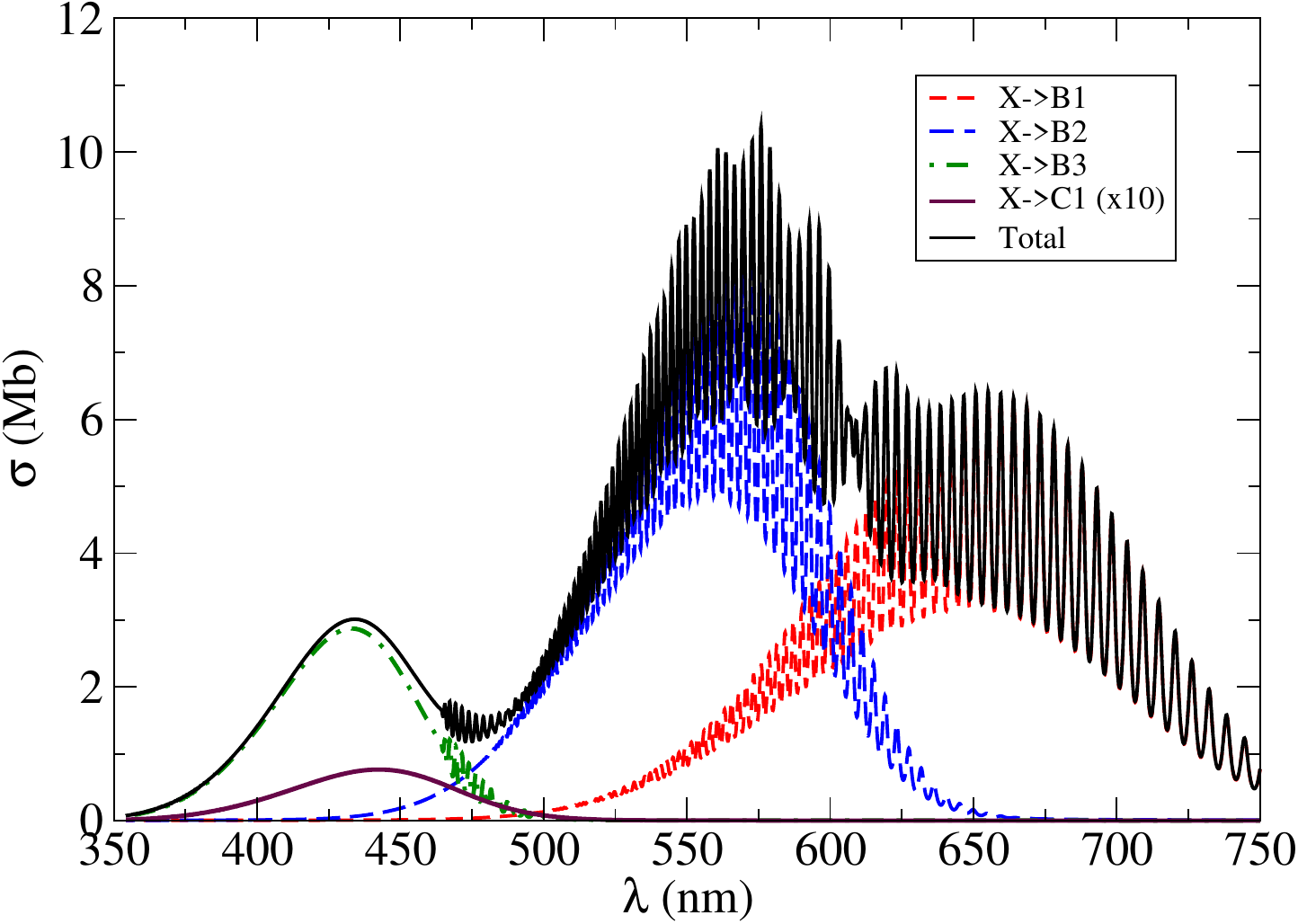}\\
\includegraphics[width=0.7\linewidth]{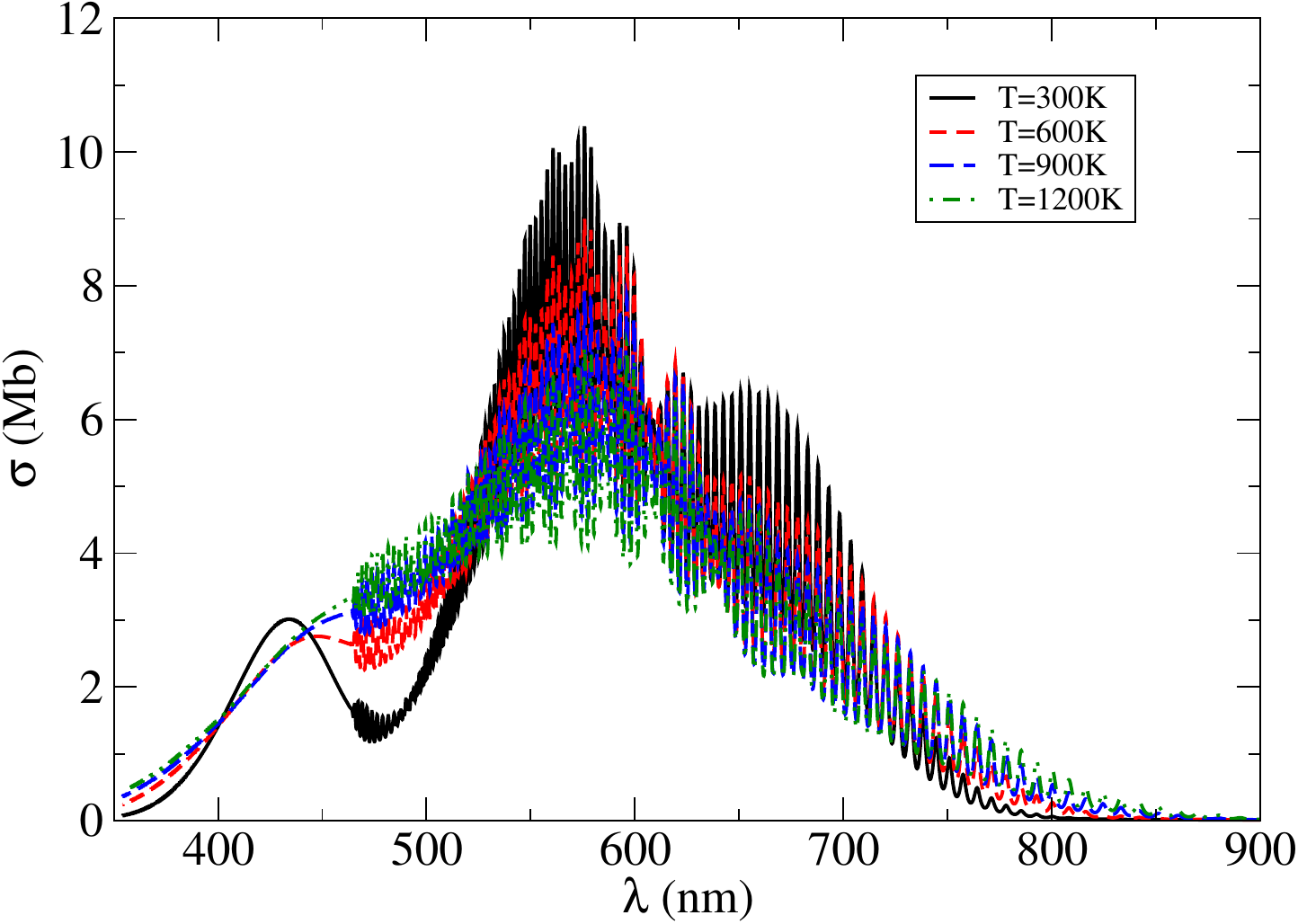}
\caption{Top: \label{fig:specti2+det} State-resolved absorption spectra of I$_2^+$ at T = 300K obtained from MRCI calculations. The spectrum corresponding to X$\rightarrow$C1 transition is multiplied by a factor of 10 to be visible. Bottom: \label{fig:specti2+T} Absorption spectra of I$_2^+$ at T = 300, 600, 900 and 1200K.}
\end{figure}

We now present in Figure~\ref{fig:specti2+det} the simulated absorption spectrum of I$_2^+$, with the breakdown of the contributions of the four absorbing states at 300K. There, we see that transitions towards states B1 and B2 form two bands partially overlapping, with vibrational substructures, and maxima at
576 nm and 655 nm. A distinct band, corresponding to the absorption towards state B3 has a maximum at 434 nm. There is no vibrational substructure for this band below 464 nm since this corresponds to sufficient excitation energies to cause the dissociation of I$_2^+$. 

As Figure~\ref{fig:specti2+T} also shows for the total absorption spectra, increasing the temperature leads to the broadening of the bands, so that already from 600K onwards there is no longer a clear separation between the bands corresponding to the absorption towards states B1, B2 and B3.

To our knowledge, the only experimental measurements of the absorption spectrum of I$_2^+$
have been carried out over limited wavelength ranges: between 720 and 770 nm by Mason~\textit{et al.}~\cite{mason_2u-x_1989} and more recently between 12065 and 13062 cm$^{-1}$ (between 766
and 829 nm) by Deng~\textit{et al.}~\cite{deng_high-resolution_2012}. In Deng~\textit{et al.}, the spectrum is vibrationally and rotationally resolved. Our calculations do not allow to obtain the rotational substructure of the spectrum but the envelope of the spectrum of Deng~\textit{et al.} are well reproduced by our spectrum: \new{the spectrum of Deng~\textit{et al.} exhibit three bands centered at around 12750, 12875 and 13000 cm$^{-1}$ (see Fig.~1 of~\cite{deng_high-resolution_2012}). In this range, our calculated spectrum shows also three bands at 12725, 12850 and 12970 cm$^{-1}$.} Note that the temperature of the experiment is not indicated in Deng~\textit{et al.} but, since the latter does not modify the wavelength position of the bands, its knowledge is not necessary for the comparison.
Note also that very recently Zhao~\textit{et al.}~\cite{zhao_spectroscopic_2024} have also calculated the spectrum of I$_2^+$
between 767 and 788 nm and reproduce with good agreement the rovibrational structure of the spectrum of Deng~\textit{et al.} 

\subsection{I$_2^-$}

\begin{figure}
\centering
\includegraphics[width=0.7\linewidth]{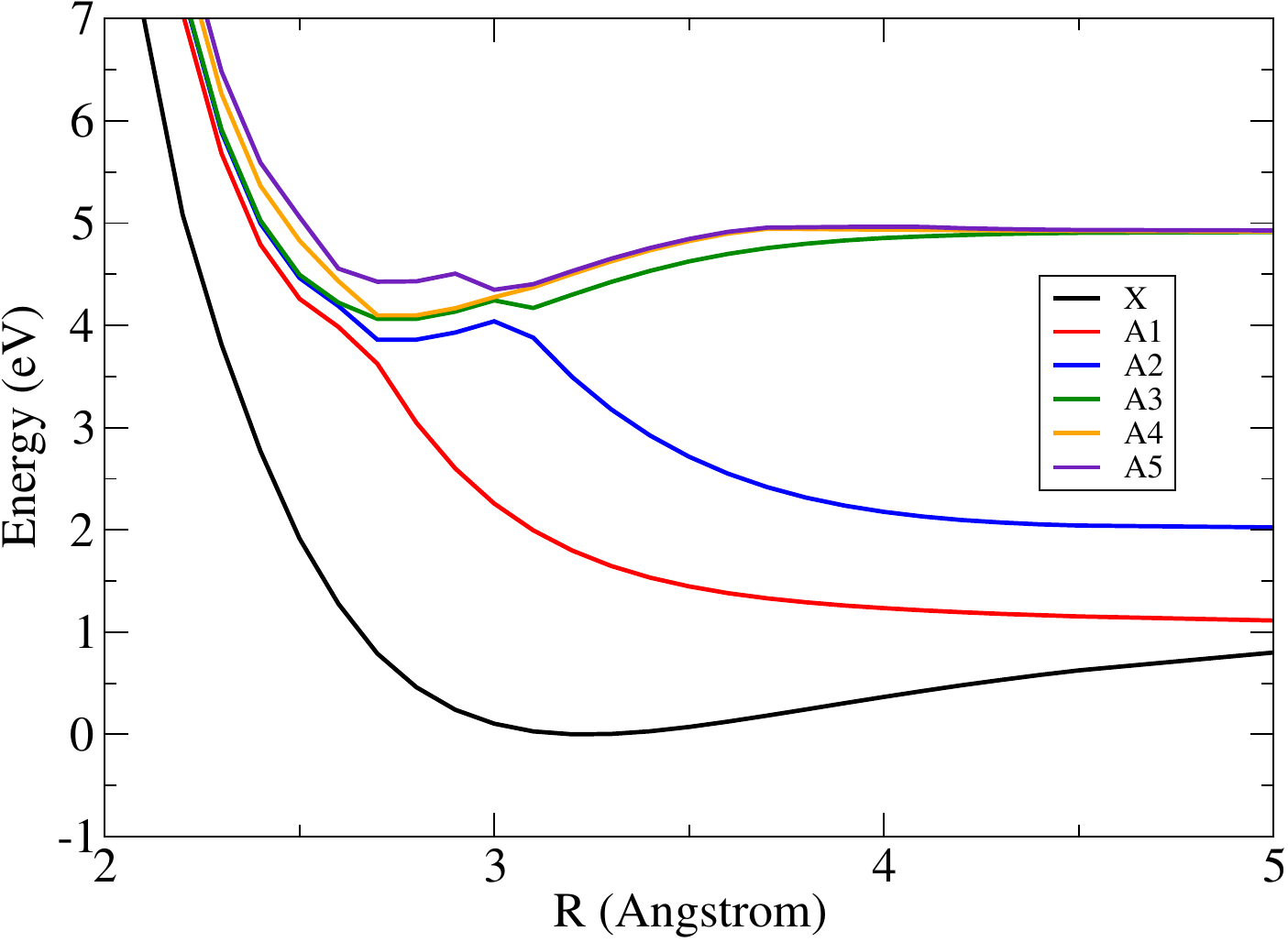}\\
\includegraphics[width=0.7\linewidth]{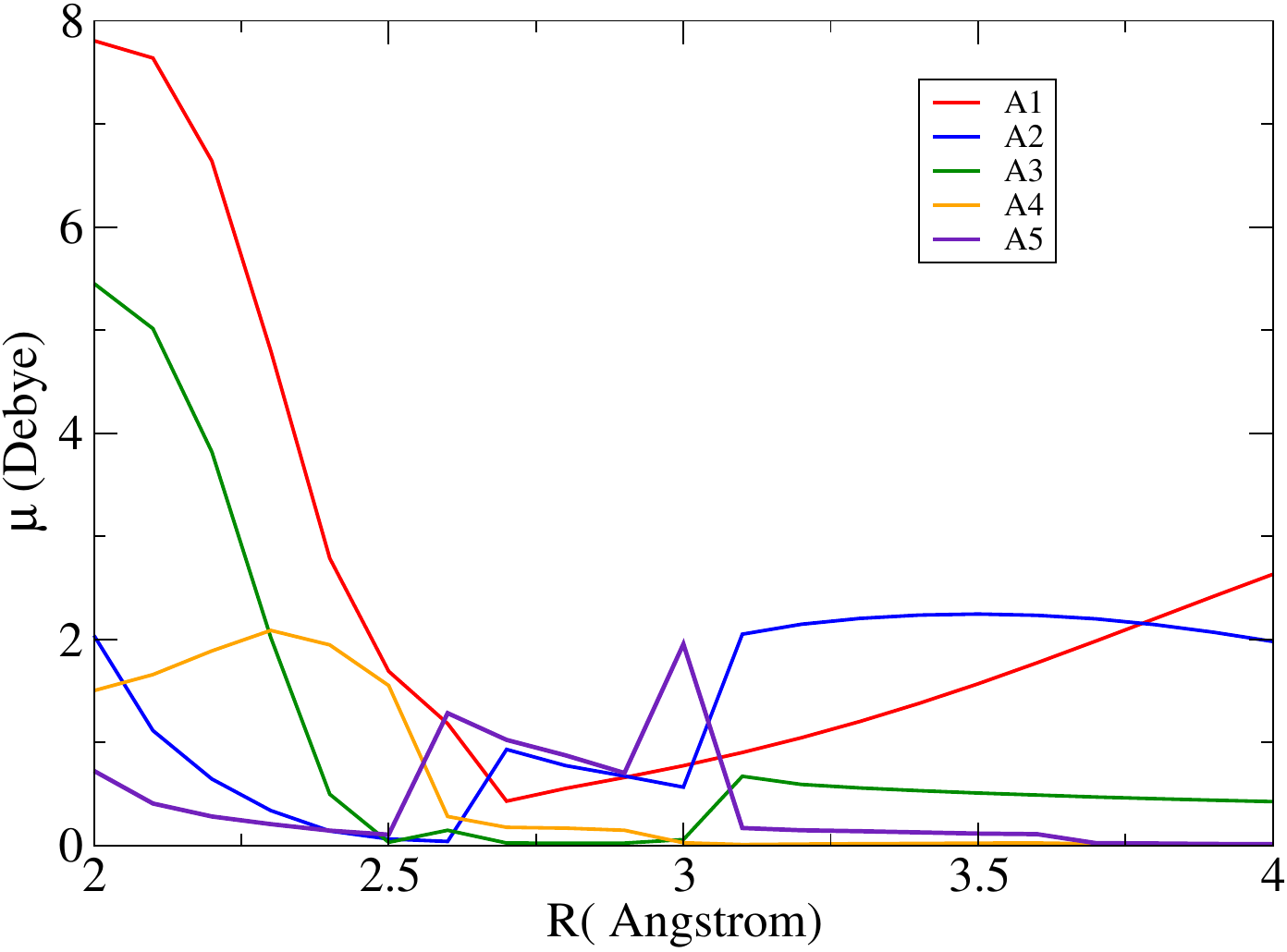}
\caption{Top: I$_2^-$ MRCI\label{fig:peci2-} Adiabatic potential energy curves (in eV) for different internuclear distances, denoted as X (black), A1 (red), A2 (blue), A3 (green), A4 (orange) and A5 (purple). Bottom:\label{fig:tdmi2-} Transition dipole moments between the ground-state and excited states. Only the relevant states for the absorption spectra are shown. }
\end{figure}

For I$_2^-$, we have computed 24 electronic states (see Table~\ref{tab:Spectro_I2-}): 5 for the symmetries ($\Omega$ = 1/2$_g$) and ($\Omega$ = 1/2$_u$), 4 for the
symmetries ($\Omega$ = 3/2$_g$) and ($\Omega$ = 3/2$_u$), and 3 for the symmetries ($\Omega$ = 5/2$_g$) and ($\Omega$ = 5/2$_u$).  The ground state of of I$_2^-$ has the symmetry of ($\Omega$ = 1/2$_u$). According to the selection rules only transitions to states with ($\Omega$ = 1/2$_g$) and ($\Omega$ = 3/2$_g$) symmetry are allowed. However, the dipole transition moments between the ground state and the ($\Omega$ = 3/2$_g$) states are negligible. Therefore, we only consider transitions from the ground state to the five lowest-lying states with ($\Omega$ = 1/2$_g$) symmetry. The corresponding potential energy curves are given in Fig.~\ref{fig:peci2-}. \\

\begin{table}
\begin{tabular}{cccccccc}
Symmetry & State & Dissociation limit  & $\Delta_{e}$ (eV) & $\omega_{e}$(cm$^{-1}$) & $\omega_{e}x_{e}$(cm$^{-1}$) & $R_{e}$(Å) & $D_{e}$(eV) \tabularnewline
\hline 
\multirow{5}{*}{$\left(\Omega=1/2\right)_{g}$} & \multirow{1}{*}{1} & \multirow{1}{*}{$^{2}\ce{P}_{3/2}+{}^{1}\ce{S}_{0}$} & 1.80 & - & - & - & -\tabularnewline
 & 2 & $^{2}\ce{P}_{1/2}+{}^{1}\ce{S}_{0}$ & 3.49 & - & - & - & -\tabularnewline
 & 3 & $^{2}\ce{P}_{3/2}+\ce{I}^{-*}$ & 4.30 & {*} & {*} & {*} & {*}\tabularnewline
 & 4 & $^{2}\ce{P}_{3/2}+\ce{I}^{-*}$ & 4.50 & {*} & {*} & {*} & {*}\tabularnewline
 & 5 & $^{2}\ce{P}_{3/2}+\ce{I}^{-*}$ & 4.53 & {*} & {*} & {*} & {*}\tabularnewline
\hline 
\multirow{6}{*}{$\left(\Omega=1/2\right)_{u}$} & \multirow{2}{*}{1} & \multirow{2}{*}{$^{2}\ce{P}_{3/2}+{}^{1}\ce{S}_{0}$} & 0 & 116 & 0.391 & 3.2 & 1.06\tabularnewline
 &  &  &  & (110 )  & (0.370 )  & (3.21) & (1.01 )\tabularnewline
 & 2 & $^{2}\ce{P}_{1/2}+{}^{1}\ce{S}_{0}$ & 2.87 & - & - & - & -\tabularnewline
 & 3 & $^{2}\ce{P}_{3/2}+\ce{I}^{-*}$ & 4.33 & {*} & {*} & {*} & {*}\tabularnewline
 & 4 & $^{2}\ce{P}_{3/2}+\ce{I}^{-*}$ & 4.40 & {*} & {*} & {*} & {*}\tabularnewline
 & 5 & $^{2}\ce{P}_{3/2}+\ce{I}^{-*}$ & 4.71 & {*} & {*} & {*} & {*}\tabularnewline
\hline 
\multirow{4}{*}{$\left(\Omega=3/2\right)_{g}$} & \multirow{1}{*}{1} & \multirow{1}{*}{$^{2}\ce{P}_{3/2}+{}^{1}\ce{S}_{0}$} & 1.26 & 52.9 & 1.88 & 3.9 & 0.21\tabularnewline
 & 2 & $^{2}\ce{P}_{3/2}+\ce{I}^{-*}$ & 4.50 & {*} & {*} & {*} & {*}\tabularnewline
 & 3 & $^{2}\ce{P}_{3/2}+\ce{I}^{-*}$ & 4.87 & {*} & {*} & {*} & {*}\tabularnewline
 & 4 & $^{2}\ce{P}_{3/2}+\ce{I}^{-*}$ & 4.93 & {*} & {*} & {*} & {*}\tabularnewline
\hline 
\multirow{4}{*}{$\left(\Omega=3/2\right)_{u}$} & \multirow{1}{*}{1} & \multirow{1}{*}{$^{2}\ce{P}_{3/2}+{}^{1}\ce{S}_{0}$} & 2.20 & - & - & - & -\tabularnewline
 & 2 & $^{2}\ce{P}_{3/2}+\ce{I}^{-*}$ & 4.41 & {*} & {*} & {*} & {*}\tabularnewline
 & 3 & $^{2}\ce{P}_{3/2}+\ce{I}^{-*}$ & 4.82 & {*} & {*} & {*} & {*}\tabularnewline
 & 4 & $^{2}\ce{P}_{3/2}+\ce{I}^{-*}$ & 4.94 & {*} & {*} & {*} & {*}\tabularnewline
\hline 
\multirow{3}{*}{$\left(\Omega=5/2\right)_{g}$} & 1 & $^{2}\ce{P}_{3/2}+\ce{I}^{-*}$ & 4.86 & {*} & {*} & {*} & {*}\tabularnewline
 & 2 & $^{2}\ce{P}_{3/2}+\ce{I}^{-*}$ & 4.93 & {*} & {*} & {*} & {*}\tabularnewline
 & 3 & $^{2}\ce{P}_{3/2}+\ce{I}^{-*}$ & 5.04 & {*} & {*} & {*} & {*}\tabularnewline
\hline 
\multirow{3}{*}{$\left(\Omega=5/2\right)_{u}$} & 1 & $^{2}\ce{P}_{3/2}+\ce{I}^{-*}$ & 4.81 & {*} & {*} & {*} & {*}\tabularnewline
 & 2 & $^{2}\ce{P}_{3/2}+\ce{I}^{-*}$ & 5.01 & {*} & {*} & {*} & {*}\tabularnewline
 & 3 & $^{2}\ce{P}_{3/2}+\ce{I}^{-*}$ & 5.03 & {*} & {*} & {*} & {*}\tabularnewline
\hline 
\end{tabular}
\caption{Spectroscopic constants for $\ce{I}_{2}^{-}$. Values in parenthesis are from~\cite{zanni_characterization_1997}}.\label{tab:Spectro_I2-}

\end{table}

The zero of energies is fixed to the dissociation limit I($^2 P_{3/2}$) + I$^-$($^1S_0$) corresponding to their electronic ground states. A second limit of dissociation is I($^2 P_{1/2}$) + I$^-$($^1S_0$) which is located at 0.94 eV above the first. Finally a third limit of dissociation corresponds to I($^2 P_{3/2}$)  and the anion I$^-$ with an electron excited to a 6s orbital, which is denoted I($^2 P_{3/2}$) + I$^{-*}$. According to the MRCI calculations this limit is located at 3.83 eV above the fundamental dissociation limit. Among the correlating states with the two lowest dissociation limits in energy only the ground state and the first state of symmetry ($\Omega$ = 3/2$_g$)  are bounded, the others are dissociative.

The states correlating with the limit I($^2 P_{3/2}$) + I$^{-*}$ present numerous avoided crossings between 2.5 and 4 \AA. The spectroscopic parameters of the ground state are generally in good agreement with those determined by Zanni~\textit{et al.}~\cite{zanni_characterization_1997} using  photoelectron spectroscopy: the difference between the  experimental and theoretical equilibrium distance is less than 0.1 \AA~while the depth of the well and the harmonicity constant are slightly overestimated in the MRCI calculations (from respectively 0.05 eV and 6 cm$^{-1}$).

As mentioned above, the dipole transition moments between the ground state and the states of symmetry $\Omega$ = 3/2$_g$ are negligible. We then only consider the transitions from the ground state to the first five states of symmetry $\Omega$ = 1/2$_g$ (denoted A1, A2, A3, A4 and A5). The transition dipole moments between the ground state and these 5 states are given in Fig.~\ref{fig:tdmi2-}.

To conclude this section, we discuss the absorption spectrum of I$_2^-$. To our knowledge, these data are the only ones concerning the absorption of I$_2^-$ in the visible range. We note the presence of avoided crossings between the A2 state and the A3, A4 and A5 states, around 2.3 \AA, i.e. close to the Franck-Condon zone. An explicit treatment of these avoided crossings would require knowledge of non-adiabatic couplings, the calculation of which is not implemented in the DIRAC MRCI code. However, we study the impact of these crossings by calculating the absorption spectra in two limiting cases. 

\begin{figure}
\centering
\includegraphics[width=0.7\linewidth]{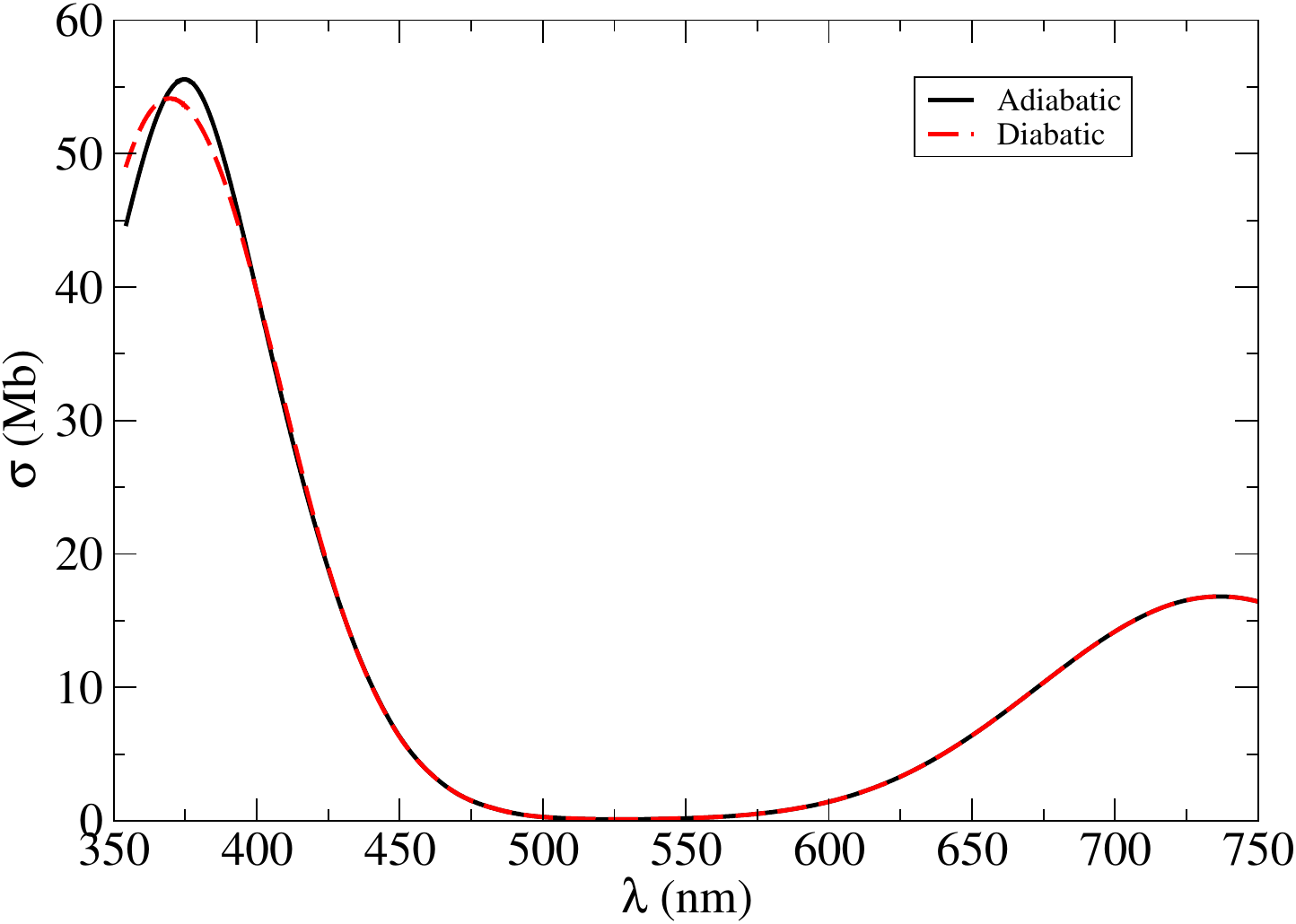}\\
\includegraphics[width=0.7\linewidth]{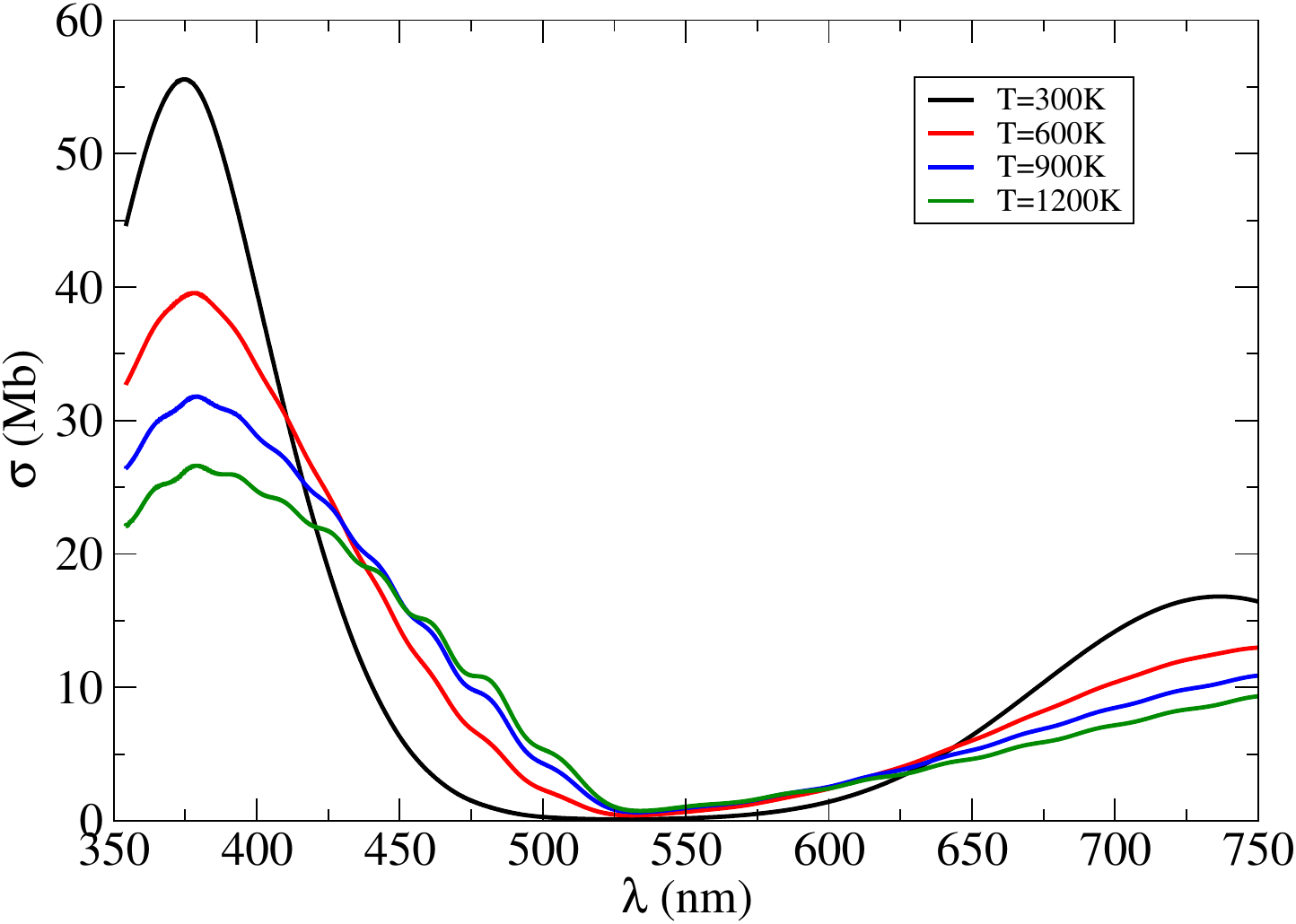}
\caption{Top:\label{fig:specti2-adiab} Absorption spectrum at T = 300K for I$_2^-$ using the adiabatic and diabatic potential energy curves. Bottom: \label{fig:specti2-} Absorption spectra at T = 300, 600, 900 and 1200K.}
\end{figure}

In the first case, the adiabatic curves are used for the absorbing states and in the second case, the diabatic extension of these curves is used. The spectra obtained in these two limiting cases, at 300K, are shown in
Figure~\ref{fig:specti2-adiab} and do not show any difference for $\lambda$ $>$ 400 nm. A band is observed
centered around 737 nm, corresponding to the transition from the ground state to the
A1 state. The other band corresponds to the transition to the A2 state. Its absorption maximum
is located at 375 nm if we consider the adiabatic potential energy curves or 380
nm if we consider the diabatic extension. Given the minimal difference between
these two spectra, we conclude that the influence of these avoided crossings on the spectrum of
I$_2^-$ is not significant. There is no vibrational structure on these bands, contrary to what is
observed for I$_2^+$, since states A1 and A2 are dissociative states. The contributions of the transitions to states A3, A4 and A5 are negligible. 

In Figure~\ref{fig:specti2-} we also present the evolution of the absorption spectrum as a function of temperature. The increase in temperature leads to the spectral broadening of the bands and to the shift of their maxima towards long wavelengths (particularly for the band corresponding to absorption towards A1).

\section{Conclusions}\label{sec.concl}

In this work we present the potential energy curves, dipole transition moments
and visible absorption spectra of I$_2$, I$_2^+$ and I$_2^-$ obtained from 4-component multireference CI calculations. The absorption spectra of I$_2^+$ and I$_2^-$ are, to our knowledge, the only ones reported over the entire visible range. Of the three spectra, that of I$_2^-$ presents the largest cross sections (up to 5 times larger than for I$_2^+$ and 10 times higher than for I$_2$) with two well-defined bands, centered around 737 and 375 nm. Its spectrum is very different from those of I$_2$ and I$_2^+$ which absorb mainly around 600 nm with a
more intense absorption and a wider wavelength range for I$_2^+$ than for I$_2$ . These spectra can be used to quantify the concentration of the species I$_2^+$ and I$_2^-$ in iodine plasmas. 

\section{Acknowledgments}
A.S.P.G acknowledges support from the Franco-German project CompRIXS (Agence nationale de la recherche ANR-19-CE29-0019, Deutsche Forschungsgemeinschaft JA 2329/6-1), PIA ANR project CaPPA (ANR-11-LABX-0005-01), I-SITE ULNE projects OVERSEE and MESONM International Associated Laboratory (LAI) (ANR-16-IDEX-0004), the French Ministry of Higher Education and Research, region Hauts de France council and European Regional Development Fund (ERDF) project CPER CLIMIBIO. S.B. thanks the École normale supérieure Paris-Saclay for the PhD scholarship.

\section{References}

\bibliography{biblio_these.bib}

\end{document}